\documentclass[showpacs, oneside, twocolumn, prl, amsmath, amssymb, 
nofootinbib, superscriptaddress]{revtex4-1}
 \pdfoutput=1
\usepackage{cases}
\usepackage{amsmath}
\usepackage{amssymb}
\usepackage{amsfonts}
\usepackage{amssymb}
\usepackage{dcolumn}
\usepackage{bm}
\usepackage{bbm}
\usepackage{graphicx}
\usepackage{xcolor}
\usepackage{array}
\usepackage{subfigure}
\usepackage{hyperref}
\usepackage{wasysym}

\newcommand{\be}{\begin{equation}}
\newcommand{\ee}{\end{equation}}
\newcommand{\ba}{\begin{eqnarray}}
\newcommand{\ea}{\end{eqnarray}}

\newcommand{\gsim}{\mathrel{\hbox{\rlap{\lower.55ex \hbox {$\sim$}}
                   \kern-.3em \raise.4ex \hbox{$>$}}}}
\newcommand{\lsim}{\mathrel{\hbox{\rlap{\lower.55ex \hbox {$\sim$}}
                   \kern-.3em \raise.4ex \hbox{$<$}}}}

\hypersetup{colorlinks=true,
            breaklinks=true,
            pdfstartview=Fit,
            linkcolor=blue,
            citecolor=blue,
            urlcolor=blue}

\begin{document}
\title{First evidence that non-metricity f(Q) gravity could challenge $\Lambda$CDM}
 
\author{Fotios K. Anagnostopoulos }
\affiliation{Department of Physics, National \& Kapodistrian University of 
Athens, 
Zografou Campus GR 157 73, Athens, Greece}

\author{Spyros Basilakos}
\affiliation{National Observatory of Athens, Lofos Nymfon, 11852 Athens,
Greece}
\affiliation{Academy of Athens, Research Center for Astronomy and
Applied Mathematics, Soranou Efesiou 4, 11527, Athens, Greece}

\author{Emmanuel N. Saridakis}
\affiliation{National Observatory of Athens, Lofos Nymfon, 11852 Athens,
Greece}
\affiliation{CAS Key Laboratory for Researches in Galaxies and Cosmology,
Department of Astronomy, University of Science and Technology of China, Hefei,
Anhui 230026, P.R. China}
\affiliation{School of Astronomy, School of Physical Sciences,
University of Science and Technology of China, Hefei 230026, P.R. China}

\begin{abstract}
We propose a novel model in the framework of $f(Q)$ gravity, which is a 
gravitational modification class arising from the incorporation of 
non-metricity. The model has General Relativity as a particular limit, it has
the same number 
of free parameters to those of $\Lambda$CDM, however at a cosmological 
framework it gives rise to a scenario 
that does not have  $\Lambda$CDM as a limit.
Nevertheless, confrontation with observations at both background and perturbation 
levels, namely with Supernovae type Ia  (SNIa), Baryonic Acoustic Oscillations 
(BAO), cosmic chronometers (CC), and  Redshift Space Distortion (RSD)  data, 
reveals that the scenario, according to AIC,  BIC and DIC information criteria, 
is in some datasets  slightly  preferred comparing to $\Lambda$CDM 
cosmology,  although in all cases the two models are statistically 
indiscriminate. 
Finally, the 
model does not exhibit early dark energy features, and thus it 
immediately passes BBN constraints, while the variation of the effective 
Newton's constant lies well inside the observational bounds. 
\end{abstract}
\maketitle

\maketitle


{\it Introduction} 
 \vspace{0.25cm}

Although General Relativity (GR) is the well-established theory for the 
description 
of the gravitational interaction, there are two main motivations that justify 
the large amount of research devoted to its modification and extension.
 The 
first arises from cosmological grounds, since modified gravity is very 
efficient in describing the universe's two phases 
of accelerated expansion 
\cite{Capozziello:2011et,CANTATA:2021ktz}. The second, and 
chronologically older, motivation  is purely theoretical, and aims towards the 
improvement of the renormalizability of General Relativity with the further 
goal to finally reach to a quantum gravitational theory  
\cite{Stelle:1976gc}. Hence, the goal is to 
construct gravitational theories that possess General Relativity as a 
particular limit, but which in general include extra degree(s) of freedom that 
are able to fulfill the above requirements. \vspace{0.16cm}
 
In this Letter we propose a specific model in the framework of the recently 
constructed $f(Q)$ modified gravity, which as we show is very efficient in 
fitting the cosmological data. 
In this class of modification, one starts from the so-called symmetric 
teleparallel theories, which is an equivalent description of gravity using the  
\emph{non-metricity} scalar $Q$  \cite{BeltranJimenez:2017tkd}, and extends it 
to an arbitrary function $f(Q)$. $f(Q)$ gravity leads to interesting 
applications
\cite{Jimenez:2019ovq, Dialektopoulos:2019mtr,
Barros:2020bgg, Bajardi:2020fxh,
Ayuso:2020dcu, Flathmann:2020zyj,DAmbrosio:2020nev,
Frusciante:2021sio,Khyllep:2021pcu}, and  trivially passes the   constraints 
arising from  gravitational wave observations
\cite{Soudi:2018dhv}.  
By confronting our new model with   data from 
Supernovae type Ia  (SNIa), Baryonic Acoustic Oscillations (BAO), Hubble 
parameter cosmic chronometers (CC) and  Redshift Space Distortions (RSD) 
f$\sigma_8$ observations,
we deduce 
that the scenario at hand     may be,  in some cases, slightly 
statistically preferred   than
$\Lambda$CDM, although it does not include it as a particular limit. 
\vspace{0.2cm}

{\it A novel model in  $f(Q)$ gravity} \vspace{0.2cm}

The action of non-metricity-based modified gravity is  
\cite{BeltranJimenez:2017tkd}  
 \begin{equation}  
 \label{fQaction}
 S = -\frac{1}{16\pi G} \int {\mathrm{d}}^4 x \sqrt{-g}  f(Q) ,
\end{equation}
where the nonmetricity scalar
\begin{equation}
\label{NontyScalar}
Q=-\frac{1}{4}Q_{\alpha \beta \gamma}Q^{\alpha \beta 
\gamma}+\frac{1}{2}Q_{\alpha \beta \gamma}Q^{ \gamma \beta 
\alpha}+\frac{1}{4}Q_{\alpha}Q^{\alpha}-\frac{1}{2}Q_{\alpha}\tilde{Q}^{\alpha} 
\,,
\end{equation}
with
$Q_{\alpha}\equiv Q_{\alpha \ \mu}^{\ \: \mu} \,   $ and
$\tilde{Q}^{\alpha} 
\equiv Q_{\mu }^{\ \: \mu \alpha}  \, ,
$ arises from contractions of the  non-metricity tensor $
    Q_{\alpha\mu\nu}\equiv\nabla_\alpha g_{\mu\nu}$. 
From these it is apparent that Symmetric Teleparallel Equivalent of General 
Relativity (and thus General Relativity) is 
obtained in the case of $f(Q)=Q$.

Varying the total action $S+S_m$, with $S_m$ the matter sector action,
one obtains the field equations as
\cite{Jimenez:2019ovq, Dialektopoulos:2019mtr}: 
\begin{eqnarray}
&&  
\frac{2}{\sqrt{-g}} \nabla_{\alpha}\left\{\sqrt{-g} g_{\beta \nu} f_{Q} 
\left[- \frac{1}{2} L^{\alpha \mu \beta}+ \frac{1}{4} g^{\mu \beta} 
\left(Q^\alpha -  \tilde{Q}^\alpha \right) \right.\right.\nonumber\\
&&\left.\left. \ \ \ \ \ \ \ \ \ \ \ \ \ \ \ \ \ \  \ \ \ \ \ \ \ \ \ \ \ \, 
- \frac{1}{8} 
\left(g^{\alpha \mu} Q^\beta + g^{\alpha \beta} Q^\mu  
\right)\right]\right\}
 \nonumber \\
&& + f_{Q} \left[- \frac{1}{2} L^{\mu \alpha \beta}- \frac{1}{8} \left(g^{\mu 
\alpha} Q^\beta 
+ g^{\mu \beta} Q^\alpha  \right)
 \right. \nonumber\\
&&\left. \ \ \ \ \ \ \ + \frac{1}{4} g^{\alpha 
\beta} \left(Q^\mu -  \tilde{Q}^\mu \right)
\right] Q_{\nu \alpha 
\beta} +\frac{1}{2} \delta_{\nu}^{\mu} f=T_{\,\,\,\nu}^{\mu}\,,
\label{eoms}
\end{eqnarray}
with 
$
L^{\alpha}_{\,\,\mu\nu}=\frac{1}{2}Q^{\alpha}_{\,\,\mu\nu}-Q^{\,\,\,\alpha}_{
(\mu\,\,\,\nu)} $  the  disformation tensor, $T_{\mu\nu}$ the energy-momentum 
tensor, and
$f_{Q}\equiv\partial f/\partial Q$. 
In order to apply it in 
a cosmological  framework we impose a flat Friedmann-Robertson-Walker 
(FRW) metric of the form 
$
ds^{2}=-dt^{2}+a^{2}(t)\delta_{ij}dx^{i}dx^{j}$, in which case equations 
(\ref{eoms}) give rise to  the two Friedmann equations \cite{Jimenez:2019ovq}
\begin{eqnarray}
\label{Fr1}
6f_QH^2-\frac12f&=& 8\pi G(\rho_m+\rho_r) \label{eqFrid},  \\
\big(12H^2f_{QQ}+f_Q\big)\dot{H}&=&-4\pi G(\rho_m+p_m+\rho_r+p_r)\,,
\label{Fr2}
\end{eqnarray}
where $H\equiv\dot{a}/a$ is  the  the Hubble function and 
with  $\rho_m$, \ $\rho_r$ and  $p_m$,\ $p_r$ the energy densities and pressures 
of 
the matter and radiation perfect fluids. Additionally, note that 
the nonmetricity scalar $Q$ in an FRW background becomes $
Q=6H^2$. Finally,  
the equations close by the consideration of the matter and 
radiation conservation equations 
\begin{eqnarray}
&&
\dot{\rho}_m+3H(\rho_m+p_m)=0
\nonumber\\
&&\dot{\rho}_r+3H(\rho_r+p_r)=0.
\end{eqnarray}

\begin{table*}[ht]
\tabcolsep 2.5pt
\vspace{1mm}
\begin{tabular}{cccccccc} \hline \hline
Model & $\Omega_{m0}$ & $h$ & $r_d $  & $\sigma_{8}$ & $\mathcal{M}$  & 
$\chi_{\text{min}}^{2} $&  $\chi_{\text{min}}/dof$  \vspace{0.05cm}\\ 
\hline
\hline
\\
 \multicolumn{7}{c}{ {\text{SNIa}/\text{CC}}}\\ 
 $Q e^{\lambda \frac{Q_0}{Q}}$ &$  0.349 \pm 0.021   $ & $  0.6828 
_{-0.0201}^{+0.0203}  $ & $-$ & $-$ & $   -19.412 \pm 0.062     $ &  $1033.37$   
& $0.968$ \\
 $\Lambda$CDM & $ 0.299 \pm 0.021$ & $0.6825_{-0.0201}^{+0.0203}     $& 
$-$ & 
$-$ & $  -19.406_{-0.062}^{+0.061} $ & $1033.267 $ &  $  0.968  $  
\vspace{0.3cm}\\
 \multicolumn{7}{c}
 {{\text{SNIa}/\text{CC}/\text{BAOs}}}\\
 \vspace{0.05cm}
$Q e^{\lambda \frac{Q_0}{Q}}$ & $ 0.353_{-0.019}^{+0.020}  $ & 
$0.6800_{-0.0199}^{+0.0201}   $ 
& 
$  148.747_{-4.144}^{+4.385}   $ & $-$ &$ 
  -19.419_{-0.062}^{+0.066}   $
& $ 1035.613   $ & $ 0.966$  \\ 

 $\Lambda$CDM & $0.304 \pm 0.0202  $ & $0.6794 \pm 0.0199 $& 
$ 148.141_{-4.112}^{+4.350}  $ & $-$ &$ 
 -19.413_{-0.062}^{+0.060}  $
& $1035.957  $ & $ 0.966$  

   \vspace{0.3cm}\\  
  \multicolumn{7}{c}{ {\text{SNIa}/\text{CC}/\text{BAOs}}}\\
  \vspace{0.05cm}
   $Q e^{\lambda \frac{Q_0}{Q}}$ & $0.339 \pm 0.020 $ & $ 
0.6864_{-0.0202}^{+0.0204}  $& 
$-$ &
$  0.703 \pm 0.0292    $ & $-19.405_{-0.062}^{+0.061}  $ & $ 1048.392 $ &  $ 
0.964 
$ \\
   $\Lambda$CDM & $  0.292 \pm 0.020$ & $0.6852_{-0.0201}^{+0.0202} $& 
$-$ & 
$0.742_{-0.031}^{+0.032}$ & $  -19.400_{-0.062}^{+0.060} $& $ 1046.640 $ &  
$0.963$  \vspace{0.1cm}
  \\  
\hline\hline
\end{tabular}
\caption[]{Observational constraints and the
corresponding $\chi^{2}_{\text {min}}$ 
for the new     $f(Q)$ gravity model 
(\ref{BGeqsfQ}), where 
``dof'' stands for degrees of freedom (defined as the number of the used data 
points minus the number of fitted parameters). In order to allow direct 
comparison  the 
concordance $\Lambda$CDM model is also included. 
 }
\label{tab:Results1}
\end{table*}

Proceeding to the perturbation level, elaborating the full perturbation 
equations \cite{Jimenez:2019ovq} one can extract the evolution equation for the 
matter 
overdensity, $\delta \equiv \delta\rho_{m}/\rho_{m}$, at sub-horizon scales 
in terms of the scale factor as 
\begin{equation}
    \label{chap2:f(T)_delta_last}
    \delta_{m}'' + \left(\frac{H'}{H}+\frac{3}{a}\right)\delta_{m}' = \frac{3 
\Omega_{m0}}{2H^2 a^5}\frac{G_{\text{eff}}}{G}\delta_{m},
\end{equation}
with primes denoting derivatives with respect to the scale factor, and 
where we have introduced the density parameters through 
$\Omega_{i}\equiv\frac{8\pi G\rho_i}{3H^2}$, with the subscript ``0'' denoting 
the   value at present time. Finally, $G_{\rm eff}$ is the effective Newton's 
constant in $f(Q)$ gravity, which is given as $G_{\rm eff} \equiv   G/f_Q $ 
\cite{Jimenez:2019ovq}.

In this work we propose the following model
\begin{equation}
f(Q)=Qe^{\lambda\frac{Q_0}{Q}},
\label{BGeqsfQ}
\end{equation} 
where  $\lambda$ is the sole model free parameter, and  $Q_0=6H_0^2$ with $H_0$ 
the current value of the Hubble parameter. For   
$\lambda = 0$ GR is recovered but not  $\Lambda$CDM since the cosmological 
constant in absent, thus this model alleviates the cosmological constant 
problem. Note that in certain periods of cosmic history, as the term $Q_0/Q$ 
decreases,
our model effectively reduces to the polynomial case $f(Q) = Q^{n}$. Thus, in a 
sense, it could be thought as an encapsulation of many $f(Q)$ models, where at a 
given cosmic time a term with particular $n$ becomes dominant.
Using  \eqref{BGeqsfQ} and \eqref{Fr1}, and considering 
$w_m\equiv p_m/\rho_m=0$ and $w_r\equiv p_r/\rho_r=1/3$, the   corresponding
normalized Hubble parameter $E^2\equiv 
H^2/H_0^2$ is written as
\begin{eqnarray}
 (E^2-2\lambda) e^{\lambda/E^2}= \Omega_{m0}a^{-3}+  \Omega_{r0}a^{-4}.
\end{eqnarray} 
Applying the above equation at present,  the parameter $\lambda$ can be 
expressed as
\begin{equation}
    \lambda = 0.5 + \mathcal{W}_{0}\left( \frac{\Omega_{m0} + 
\Omega_{r0}}{2e^{1/2}}\right)\,,
    \label{lambdaeq}
\end{equation}
with $\mathcal{W}_{0}$   the principal branch of the Lambert function. Hence, 
the scenario at hand has exactly the same number of free parameters with 
$\Lambda$CDM, and as we will see this is  a key reason for its statistical 
preference over the latter. 
Lastly, the effective Newton's 
constant becomes 
\begin{eqnarray}
&& G_{\text{eff}}= \frac{G}{e^{\lambda\frac{Q_0}{Q}}\left(
1-\lambda\frac{Q_0}{Q}
\right)}.
 \label{Geffeq}
\end{eqnarray}

 \vspace{0.5cm}
 
{\it Data, Methodology and Results} \\

 We employ Bayesian analysis and the likelihood function 
$\mathcal{L}_{\textrm{tot}} \sim \textrm{exp}(-\chi^2_{\textrm{tot}}/2)$, where 
$\chi^2_{\textrm{tot}}$ is obtained from different sums of 
$\chi^2_{\textrm{SNIa}},\chi^2_{\textrm{BAOs}}, \chi^2_{\textrm{CC}}, 
\chi^2_{\textrm{RSD}}$, to be 
specified below.
The quantities $\chi^2_{\textrm{SNIa}},\ \chi^2_{\textrm{RSD}}$ and 
$\chi^2_{\textrm{CC}}$ are defined 
and described in \cite{Anagnostopoulos:2019miu}.  In contrast with the 
latter, we use the  full SNIa dataset in order to avoid biases induced 
via the binning procedure. Moreover, as explained at \cite{Kjerrgren:2021zuo}, 
we employ only a subset of the CC dataset. Since RSD data were 
extracted by imposing $\Lambda$CDM as a reference model, we use the 
simple correction factor 
described in \cite{Macaulay:2013swa}. Additionally, we utilize the BAOs 
dataset of 
\cite{BOSS:2016wmc}, as they employ 
fiducial cosmology corrections (i.e the term $r_\textrm{d}/r_{\textrm{fid}}$).
We sample the posterior distribution of the parameters applying
the MCMC method as
implemented within the open-source Python package emcee 
\cite{ForemanMackey:2012ig}, 
for the following cases:
\begin{enumerate}
    \item SNIa+CC datasets, with free parameters 
$\Omega_{m0},h,\mathcal{M}$.
    \item SNIa+CC+BAOs, with free parameters 
$\Omega_{m0},h,r_d,\mathcal{M}$.
    \item SNIa+CC+RSD, with free parameters 
$\Omega_{m0},h,\sigma_{8},\mathcal{M}$.
\end{enumerate}
Note that $\mathcal{M}$ is the intrinsic free parameter of
Pantheon dataset (see \cite{Anagnostopoulos:2019miu} and references therein) and 
we neglect $\Omega_{r0}$.
We apply the aforementioned setup for our $f(Q)$ model (\ref{BGeqsfQ}), along 
with $\Lambda$CDM
to allow for a direct comparison. In all cases  1000 walkers and 2500 states 
are 
employed along with flat priors on the parameters. Although
the particular $f(Q)$ model considered here has the same number of 
free parameters with $\Lambda$CDM, the functional form of the Hubble 
rate is different and thus the relative fitting quality could only be compared 
in 
the context of information criteria differences.

\begin{figure}[!]
\includegraphics[width=0.485\textwidth]{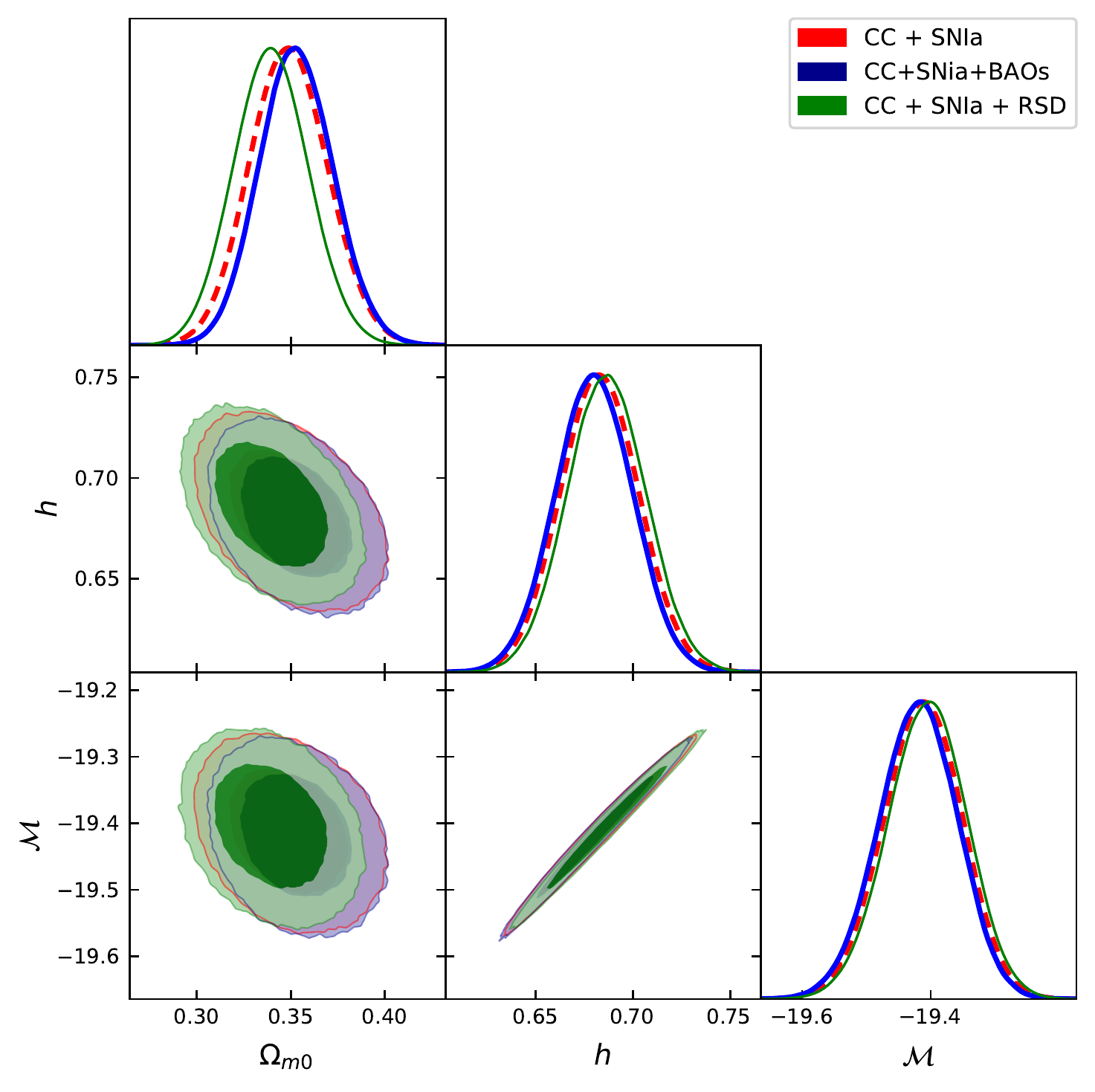}
\caption{ {\it{The $1\sigma$ and  $2\sigma$  iso-likelihood 
contours for the $f(Q)$ model 
(\ref{BGeqsfQ}), for the 2D subsets of the parameter space
$(\Omega_{m0},h,\mathcal{M})$, using graphic package getdist 
\cite{lewis2019getdist}. We have used 
joint 
analysis of various datasets (see text).}}
}
\label{fig:ESD}
\end{figure}

\begin{figure}[!]
\includegraphics[width=0.485\textwidth]{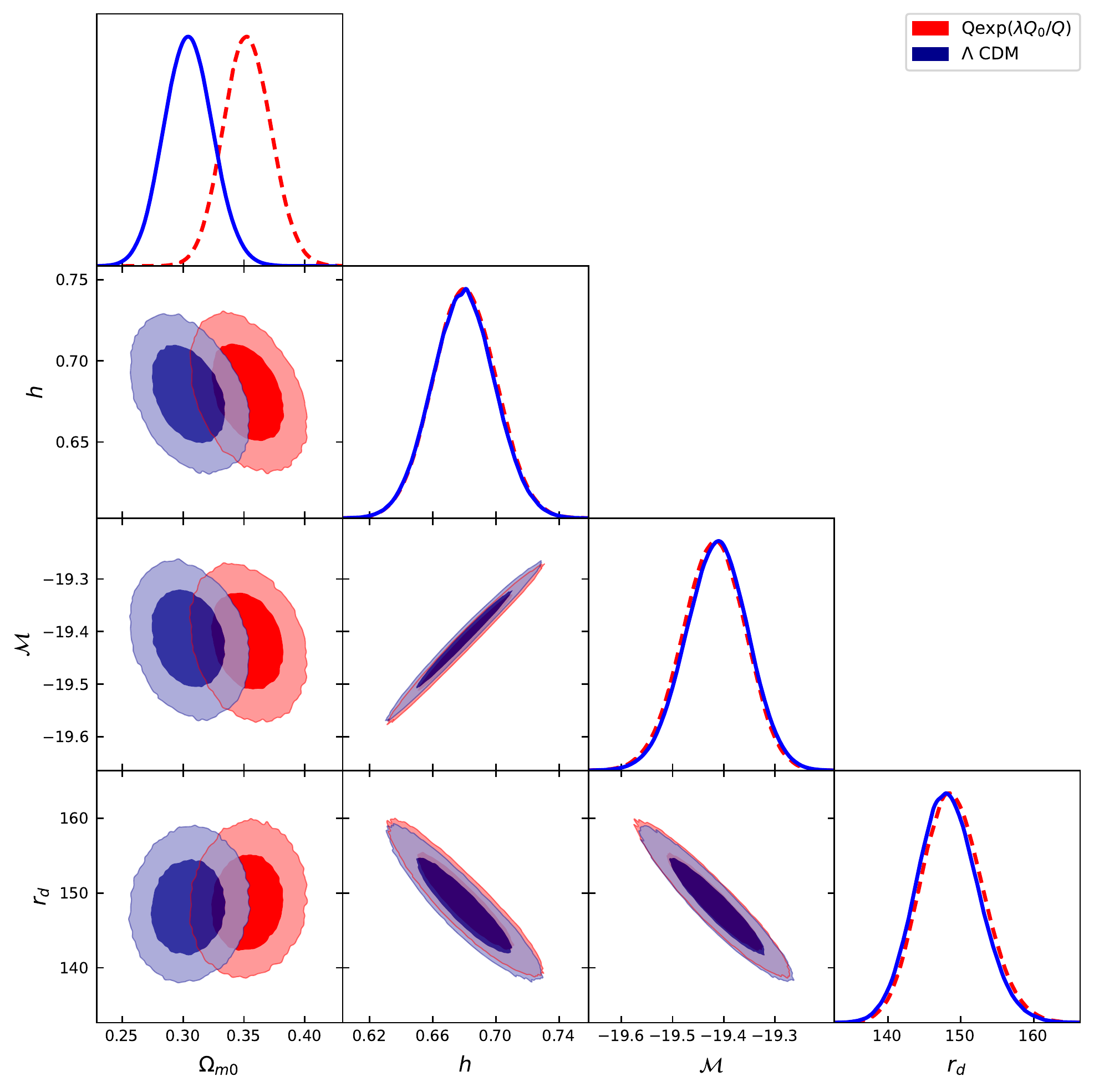}
\caption{ {\it{The $1\sigma$ and  $2\sigma$  iso-likelihood 
contours for  the $f(Q)$ model 
(\ref{BGeqsfQ}), as well as for the $\Lambda$CDM scenario, 
for the 2D subsets of the parameter space
$(\Omega_{m0},h,\mathcal{M},r_d)$, using graphic package getdist 
\cite{lewis2019getdist}.  We have used the
joint 
analysis of SNIa+CC+BAOs datasets (see text).}}
}
\label{fig:trigBothModels}
\end{figure}

We present the results on the parameters in Table \ref{tab:Results1}.  
Additionally, in Fig. \ref{fig:ESD} we show the corresponding  contour plots for 
the extracted model parameters for all considered datasets. Finally, in Fig. 
\ref{fig:trigBothModels} we depict
the contour plots for both  the $f(Q)$ and $\Lambda$CDM models, for the case of 
Pantheon+CC+BAOs datasets.  
Concerning the parameter values, in all cases $\Omega_{m0}$ is close to the 
Planck 
value, while the deviations from the concordance model is evident.
Moreover, as a consistency check, we compare the sound horizon at baryon drag 
epoch, $r_d$, with the model-independent one from \cite{Haridasu:2018gqm}, and 
we 
observe 1$\sigma$ compatibility. 
Note that concerning the $H_0$ and $\sigma_8$ 
tensions, in order to be able to extract safe results we should analyze the 
full CMB spectrum and compare the corresponding parameter values with the
current ones. Lastly, we mention that for the case of Pantheon+CC+BAOs dataset 
the parameter $\lambda$ is found as  $\lambda = 0.371 \pm 0.008$ at 1$\sigma$ 
confidence level.

Proceeding forward, in Table \ref{tab:Results2}  we compare the fitting quality 
of the 
present $f(Q)$ model with the corresponding one of $\Lambda$CDM cosmology, 
through the use of   the Akaike Information Criterion (AIC), the Bayesian 
Information Criterion (BIC), and the
Deviance 
Information Criterion (DIC), see \cite{Anagnostopoulos:2019miu} and references 
therein. For SNIa + CC datasets, we find that 
  $\Delta$IC is almost 0, and thus the two models are 
statistically compatible. For SNIa + BAOs + CC,  our $f(Q)$ model is slightly 
more 
preferred by the data. In contrast, for SNIa + BAOs + RSD 
the $f(Q)$ 
model is deemed inferior by the data, however still statistically 
indistinguishable from $\Lambda$CDM. We mention here
(without considering the possibility that the BAOs data  may include numerical 
relics)   that  the fact that these BAOs employ a free parameter to correct for 
the imposed cosmology, while the f$\sigma_{8}$ data do not, 
could be an indication for 
circularity problem with the RSD data. 
Hence,  although prominent, the present $f(Q)$ model,  since 
it has no $\Lambda$CDM limit,   gets ``punished''
by those datasets that incorporate  $\Lambda$CDM as a reference model.
Finally, we mention    that our model includes many models as subcases (i.e 
polynomial) at particular times within the cosmic history, however it   
maintains the same 
number of free parameters with $\Lambda$CDM cosmology.

\begin{table}[ht]
\begin{center}
\tabcolsep 2.0pt
\vspace{1mm}
\begin{tabular}{ccccccc} \hline \hline
Model & AIC & $\Delta$AIC & BIC &$\Delta$BIC & DIC & $\Delta$DIC
 \vspace{0.02cm}\\ 
 \hline
\hline
 \\
 \multicolumn{7}{c}{ {\text{SNIa}/\text{CC}}}\\ 
  $Q e^{\lambda \frac{Q_0}{Q}}$ &  $1039.390$ & $0.1$ & $1054.30$ & $0.106$ & 
$1039.320 $ 
& $0.115$ \\  
$\Lambda$CDM & $ 1039.290$ & $0$ & $1054.194$ & $0.0$ & $1039.205$ & $0$  \\ 
 \vspace{0.05cm}\\  
  \multicolumn{7}{c}{ {\text{SNIa}/\text{CC}/\text{BAOs}}}\\
  \vspace{0.05cm}
 $Q e^{\lambda \frac{Q_0}{Q}}$ & $  1043.650$ & $0$ & 
$  1063.537$ & $0$ &$ 1043.542 $
& $0$ \\ 
$\Lambda$CDM & $1043.994 $ & $0.344$ & $1063.881$ & $0.344$ & $1043.888$ & 
$0.346$\\

 \vspace{0.05cm}\\ 
  \multicolumn{7}{c}{ {\text{SNIa}/\text{CC}/\text{RSD}}}\\

   $Q e^{\lambda \frac{Q_0}{Q}}$ & $1056.430$ & $1.753$ & 
$  1076.3749 $ & $1.751$ &$ 1056.3198$
& $1.750$ \\  
$\Lambda$CDM & $1054.677$ & $0$ & $1074.624 $ & $0$ & $1054.570$ & 
$0$ \vspace{0.02cm}\\ 
 \hline\hline
\end{tabular}
\caption{The information criteria 
AIC, BIC and DIC for the examined cosmological models,
alongside the corresponding differences
$\Delta\text{IC} \equiv \text{IC} - \text{IC}_{\text{min}}$.
\label{tab:Results2}}
\end{center}
\end{table}

{\it Conclusions} \vspace{0.2cm}

 We proposed a novel $f(Q)$ model  
and we confronted it against observational data (SNIa, BAOs, CC and RSD). 
For CC + SNIa datasets the two models are statistically compatible, however for 
CC + SNIa + BAOs datasets  the $f(Q)$ model is  slightly  statistically 
preferred comparing 
to $\Lambda$CDM one. On the other hand, in the case of 
RSD + CC + SNIa data, the $\Lambda$CDM  paradigm is  slightly
preferred by the data, although the two models remain statistically equivalent.

In the large redshift limit (i.e at large $E^2(z)\equiv 
H^2(z)/H_0^2$) the proposed $f(Q)$ tends to $Q$ and thus the scenario at hand 
tends to GR, hence it trivially passes the early universe constraints and in 
particular the BBN ones. Additionally, knowing the observational bounds of 
$E^2(z)$ throughout the evolution, and using (\ref{lambdaeq}), from 
(\ref{Geffeq})
we deduce that throughout the evolution $|\frac{G_{\rm eff}}{G}-1|$ remains smaller 
than 0.1 and therefore it satisfies the observational constraints \cite{Nesseris:2017vor}.

 In summary, our results could serve as motivation for further study 
of the present model, as well as $f(Q)$ gravity in general,
as it constitutes one of the first 
alternatives to the concordance model   that apart from the 
fact that  it might be    preferred  by the data (at least by some 
datasets),  it does not face the cosmological constant problem since it 
does not include a ``hidden'' cosmological constant inside the $f(Q)$ form. 
Further studies on this model, 
using the full CMB and LSS spectra, weak lensing data and other datasets, could 
enlighten our findings and verify whether the present $f(Q)$ model
outperforms the concordance one or not. \\
 
Acknowledgments:
F.K. Anagnostopoulos wishes to thank Jiaming Shi for commenting on a draft of this paper.


\begin{thebibliography}{}


\bibitem{Capozziello:2011et}
S.~Capozziello and M.~De Laurentis,
Phys. Rept. \textbf{509}, 167-321 (2011).

\bibitem{CANTATA:2021ktz}
E.~N.~Saridakis \textit{et al.} [CANTATA],
[arXiv:2105.12582 [gr-qc]].
  
 


\bibitem{Stelle:1976gc}
  K.~S.~Stelle,
  Phys.\ Rev.\ D {\bf 16}, 953 (1977).



 

\bibitem{BeltranJimenez:2017tkd}
J.~Beltr\'an Jim\'enez, L.~Heisenberg and T.~Koivisto,
Phys. Rev. D \textbf{98}, no.4, 044048 (2018).

\bibitem{Dialektopoulos:2019mtr}
K.~F.~Dialektopoulos, T.~S.~Koivisto and S.~Capozziello,
Eur. Phys. J. C \textbf{79}, no.7, 606 (2019).


\bibitem{Barros:2020bgg}
B.~J.~Barros, T.~Barreiro, T.~Koivisto and N.~J.~Nunes,
Phys. Dark Univ. \textbf{30}, 100616 (2020).

\bibitem{Jimenez:2019ovq}
J.~Beltr\'an Jim\'enez, L.~Heisenberg, T.~S.~Koivisto and S.~Pekar,
Phys. Rev. D \textbf{101}, no.10, 103507 (2020).
 
 
 

     

  

\bibitem{Bajardi:2020fxh}
F.~Bajardi, D.~Vernieri and S.~Capozziello,
Eur. Phys. J. Plus \textbf{135}, no.11, 912 (2020).
 


\bibitem{Ayuso:2020dcu}
I.~Ayuso, R.~Lazkoz and V.~Salzano,
Phys. Rev. D \textbf{103}, no.6, 063505 (2021).


 

 

\bibitem{Flathmann:2020zyj}
K.~Flathmann and M.~Hohmann,
Phys. Rev. D \textbf{103}, no.4, 044030 (2021).



\bibitem{Frusciante:2021sio}
N.~Frusciante,
Phys. Rev. D \textbf{103}, no.4, 044021 (2021).


 
 


\bibitem{Khyllep:2021pcu}
W.~Khyllep, A.~Paliathanasis and J.~Dutta,
Phys. Rev. D \textbf{103}, no.10, 103521 (2021).

 \bibitem{DAmbrosio:2020nev}
F.~D'Ambrosio, M.~Garg and L.~Heisenberg,
Phys. Lett. B \textbf{811}, 135970 (2020).
 


\bibitem{Soudi:2018dhv}
I.~Soudi, G.~Farrugia, V.~Gakis, J.~Levi Said and E.~N.~Saridakis,
Phys. Rev. D \textbf{100}, no.4, 044008 (2019).

 


\bibitem{Anagnostopoulos:2019miu}
F.~K.~Anagnostopoulos, S.~Basilakos and E.~N.~Saridakis,
Phys. Rev. D \textbf{100}, no.8, 083517 (2019).

\bibitem{Kjerrgren:2021zuo}
A.~A.~Kjerrgren and E.~Mortsell,
[arXiv:2106.11317 [astro-ph.CO]].


\bibitem{Macaulay:2013swa}
E.~Macaulay, I.~K.~Wehus and H.~K.~Eriksen,
Phys. Rev. Lett. \textbf{111}, no.16, 161301 (2013).


\bibitem{BOSS:2016wmc}
S.~Alam \textit{et al.} [BOSS],
Mon. Not. Roy. Astron. Soc. \textbf{470} (2017) no.3, 2617-2652.


\bibitem{ForemanMackey:2012ig}
D.~Foreman-Mackey, D.~W.~Hogg, D.~Lang and J.~Goodman,
Publ. Astron. Soc. Pac. \textbf{125}, 306-312 (2013).


\bibitem{Haridasu:2018gqm}
B.~S.~Haridasu, V.~V.~Lukovi\'c, M.~Moresco and N.~Vittorio,
JCAP \textbf{10}, 015 (2018).

\bibitem{lewis2019getdist}
Lewis, A. GetDist,
{\it{Monte Carlo sample analyzer}}, 
  Astrophysics Source Code Librar.


\bibitem{Nesseris:2017vor}
S.~Nesseris, G.~Pantazis and L.~Perivolaropoulos,
Phys. Rev. D \textbf{96}, no.2, 023542 (2017).
 

\end{thebibliography}
\end{document}